\documentclass[%
 reprint,
superscriptaddress,
 amsmath,amssymb,
 aps,
prl,
]{revtex4-1}

\usepackage{graphicx}% Include figure files
\usepackage{dcolumn}% Align table columns on decimal point
\usepackage{bm}% bold math
\begin{document}

\preprint{APS/123-QED}

\title{Finding and breaking the realistic rate-distance limit of continuous variable
\\quantum key distribution}% Force line breaks with \\

\author{Xuyang Wang}
\email{wangxuyang@sxu.edu.cn}
\affiliation{State Key Laboratory of Quantum Optics and Quantum Optics Devices, Institute of Opto-Electronics, Shanxi University, Taiyuan 030006, China}
\affiliation{Collaborative Innovation Center of Extreme Optics, Shanxi University, Taiyuan 030006, China}

\author{Siyou Guo}%
\affiliation{State Key Laboratory of Quantum Optics and Quantum Optics Devices, Institute of Opto-Electronics, Shanxi University, Taiyuan 030006, China}

\author{Pu Wang}
\affiliation{State Key Laboratory of Quantum Optics and Quantum Optics Devices, Institute of Opto-Electronics, Shanxi University, Taiyuan 030006, China}

\date{\today}% It is always \today, today,
             %  but any date may be explicitly specified

\begin{abstract}
In this work, the rate-distance limit of continuous variable quantum key distribution is studied. We find that the excess noise generated on Bob's side and the method for calculating the excess noise restrict the rate-distance limit. Then, a realistic rate-distance limit is found. To break the realistic limit, a method for calculating the secret key rate using pure excess noise is proposed. The improvement in the rate-distance limit due to a higher reconciliation efficiency is analyzed. It is found that this improvement is dependent on the excess noise. From a finite-size analysis, the monotonicity of the Holevo bound versus the transmission efficiency is studied, and a tighter rate-distance limit is presented.
\end{abstract}

% PACS, the Physics and Astronomy
                             % Classification Scheme.
%\keywords{Suggested keywords}%Use showkeys class option if keyword
                              %display desired
\maketitle

%\tableofcontents

%\section{Introduction}

\textit{Introduction.}---Quantum key distribution (QKD) allows two distant parties to share secret keys based on the laws of quantum physics. Anybody who eavesdrops on
the information will be discovered, and the eavesdropped information can be
subtracted to ensure security. It is expected that this technology will
have great potential for applications in the future. Usually, QKD can be divided into discrete variable (DV) and
continuous variable (CV) domains. The protocols in either domain have their
advantages \cite{Scarani,Weedbrook1,Pirandola}.

The first protocol in the CV domain was proposed in 1999 using squeezed
states \cite{Ralph}, and early CV protocols primarily focused on squeezed and entangled states \cite{Cerf,Gottesman,Silberhorn}. In 2002, Grosshans and Grangier proposed the famous GG02 protocol \cite{Grosshans1}. This protocol enabled the realization of QKD by using only coherent states instead of nonclassical states. Then, the reverse reconciliation method was designed to beat the 3-dB limit. In 2003, the GG02 protocol was demonstrated experimentally \cite{Grosshans2}. Since then, CV QKD entered a period of rapid development. Various protocols were proposed and realized experimentally \cite{Weedbrook2,Garcia-Patron,Lodewyck1,Jouguet,XYWang1,SXLong,Usenko,CWang,BQi,DHuang1,XYWang2}. Prototypes were also realized, and some field tests were reported \cite{Fossier,DHuang2,YMLi}. At present, CV QKD can be realized using low-cost off-the-shelf components and have good compatibility with classical communication systems \cite{Karinou}. Thus, it is considered a promising candidate for enabling the deployment of quantum cryptography in future networks. However, the longest distance of CV QKD remains limited to $\sim $100 km in experiments. The reasons behind such a short distance are the reconciliation efficiency, excess noise, and finite-size effect. Usually, to minimize the finite-size effect, a typical method is increasing the total number of samples. Thus, improving the reconciliation efficiency, reducing the excess noise, and minimizing the finite-size effect are the main goals for developing methods that enhance the performance of various CV QKD systems.

Fortunately, owing to the incessant efforts of researchers worldwide, the reconciliation efficiency is continuously increased. At present, a
reconciliation efficiency of 99\%, which is close to the Shannon limit, can
be achieved \cite{ Milicevic}. In this study, it was found that the improvement in the rate-distance limit due to higher reconciliation efficiency is dependent on the excess noise. Based on a detailed investigation of the CVQKD system, we found that the excess noise generated on Bob's side and the method for calculating the excess noise restrict the rate-distance limit. Then, a realistic rate-distance limit was found. To break the realistic limit, a method for calculating the secret key rate using pure excess noise is proposed. Revisiting the finite-size effect, we find a loophole due to the monotonicity of the Holevo bound. Then, a tighter rate-distance limit was used to ensure the security of system.

%\section{The rate-distance limits and experiment results of several typical protocols}

\textit{Rate-distance limits and experiment results of several typical protocols.}---There are various protocols in CV QKD, such as one-way,
two-way, and measurement-device-independent (MDI) protocols. Here,
we mainly focus on one-way protocols that are experimentally
demonstrated well. The rate-distance limits of several typical protocols are presented
in Fig.1, and certain representative experimental results are also
presented.

% FIG. 1
\begin{figure}[tbp]
\centerline{
\includegraphics[width = 86mm]{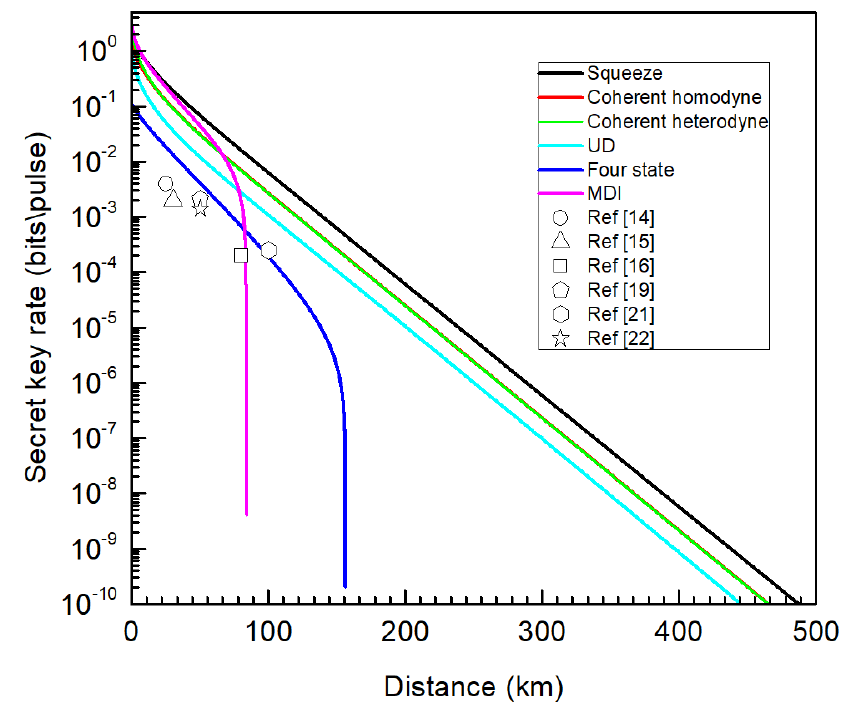}
}
\caption{Rate-distance limits and experimental results of several typical protocols.}
\label{Fig 1}
\end{figure}

In Fig. 1, we can see that the transmission distances obtained by various
experiments are less than or approximately 100 km, although the maximum
transmission distance in theory can reach almost 500 km in the asymptotic
case. To achieve reasonable rate-distance limits, the following parameters are used: reverse reconciliation efficiency $\beta =0.95$,
excess noise $\varepsilon =0.01$, detection efficiency $\eta =0.6$,
electronic noise $\upsilon _{e}=0.1$, and a transmission loss of 0.2
dB/km. The modulation variance is optimized according to the transmission
distance (or transmission efficiency $T$). It is noted that all variances
are normalized to the shot noise $N_{0}$. For the entangled-state protocol, when the sender Alice uses homodyne detection, the same rate-distance limit as that for the squeezed-state protocol will be attained. When Alice uses heterodyne detection, the rate-distance limit will be that of the coherent-state protocol. Thus, the entangled-state protocol is not presented. For the presented protocols, the squeezed-state protocol with homodyne detection performs the best (black
solid line). The coherent-state protocol with heterodyne detection (green
solid line) performs better than that with homodyne detection (red solid
line) at a short distance. With increasing distance, their rate-distance
limits nearly converge to one line. Thus, the green solid line overlaps the
red solid line. The Unidimensional (UD) coherent-state protocol has comparable performance with the normal two dimensional coherent protocols. It has the advantages of easy modulation and low costs and requires fewer random numbers \cite{XYWang2,PWang}.
The non-Gaussian state protocols have the potential to realize high-speed
gigahertz quantum communication using quaternary phase shift keying (QPSK)
technology \cite{ZQu}. The modulated states are larger, and the performance
is better. When the number of modulated states is eight, the performance approaches
the two-dimensional coherent-state protocol \cite{Leverrier1}. Here, the typical four-state protocol is selected to be presented. The hot CV MDI protocol is
also presented \cite{Pirandola2} because it becomes very sensitive to the detection
efficiency, where the detection efficiency is set to $\eta =1$ and the
excess noise is set to $\varepsilon =0.002$. To maximize the transmission
distance, the middle part Clair is on Alice's side. A longest transmission
distance of 80 km can be achieved.

%\section{Rate-distance limit improvement due to reconciliation efficiency}

\textit{Improvement in the rate-distance limit due to the reconciliation efficiency.}---From a review of the experimental results, we can see that the realistic transmission distance is limited to $\sim $100 km. One method for improving the performance is to improve the reconciliation efficiency. The reconciliation efficiency was recently improved to 0.99 \cite{Milicevic}. This value is close to the Shannon limit. In the following, we will calculate the improvement in the rate-distance limit due to this higher efficiency based on coherent homodyne protocol (the famous GG02 protocol).

% FIG. 2
\begin{figure}[tbp]
\centerline{
\includegraphics[width=86mm]{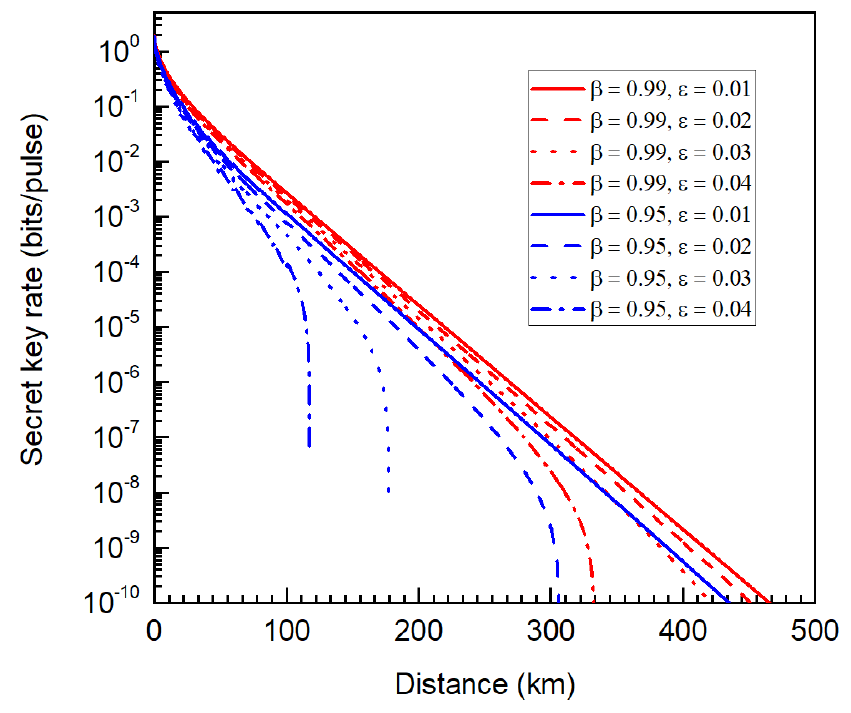}
}
\caption{ Rate-distance limits for reconciliation efficiencies of $\beta =0.95$ and $\beta =0.99$.}
\label{Fig 2}
\end{figure}

In Fig. 2, the red lines present the improved rate-distance limits due to a higher reconciliation efficiency of $\beta =0.99$ at different excess noise conditions.
Comparing them with the blue lines that represent the performance at a
lower efficiency $\beta =0.95$ with different excess noise, we observe that the improvement in the rate-distance limit is dependent on the excess noise. If the excess noise is larger, the improvement is more obvious. Because the limit of the distance at $\beta =0.95$
and excess noise of 0.01 is far larger than the realistic transmission distance, we
can infer that the reconciliation efficiency is not the primary reason limiting
the rate-distance limits when the excess noise is $\sim $0.01.

\textit{Realistic Rate-distance limit.}---Typically, there is a law that states that the excess noise is larger when the distance is higher. Considering that the excess noise $\varepsilon _{l}$ from the fiber is small \cite{Lodewyck2} and the excess noise $\varepsilon _{a}$ generated on Alice's side is constant, this law remains applicable. In the experiment, controlling the excess noise to less than 0.01 at 100 km is
extremely challenging \cite{Jouguet,DHuang1}. Through a careful
analysis of the experimental system, we find that the excess noise $\varepsilon
_{b}$\ generated on Bob's side and the method for calculating the excess noise
result in this law. The excess noise generated on Bob's side is mainly
attributed to the measurement error and the phase modulator used to switch
the bases. To illustrate this law clearly, a typical model for describing
the relation for the data in CV QKD is as follows:
\begin{equation}
y=t\cdot x+z,
\end{equation}
where the variable $y$\ represents the quadrature received by Bob,
and the variance of $y$ is usually denoted as $V_{B}$. The variable $x$\ is
the data used to modulate the state by Alice, and the variance of $x$ is
usually denoted as $V_{A}$. The variable $z$ follows a Gaussian
distribution with variance $\sigma ^{2}=N_{0}+\upsilon _{e}+\eta
T\varepsilon $ and a mean of zero. Thus, Eq. (1) can be transformed into
a variance form as

\begin{equation}
V_{B}=\eta T\left( V_{A}+\varepsilon \right) +N_{0}+\upsilon _{e}=\eta
TV_{A}+\eta T\varepsilon +N_{0}+\upsilon _{e}.
\end{equation}
The ideal excess noise $\varepsilon$ can be calculated by

\begin{equation}
\varepsilon =\left( V_{B}-\eta TV_{A}-N_{0}-\upsilon _{e}\right) /\eta
T=\varepsilon _{a}+\varepsilon _{l}+\varepsilon _{b}.
\end{equation}

Typically, $\varepsilon $ is normalized to the input
port of the channel for security. We assume that the excess noise from Alice's
setup, channel, and Bob's setup are stable. From Eq. (3), we can see that the noise will remain unchanged when the transmission efficiency varies. This contradicts the law that the excess noise is larger when the transmission distance is longer. We note the fact that $\varepsilon
_{b}$\ generated on Bob's side is not attenuated with the channel transmission efficiency as shown in the following:

\begin{equation}
V_{B}=\eta TV_{A}+\eta T\left( \varepsilon _{a}+\varepsilon _{l}\right)
+\varepsilon _{b}+N_{0}+\upsilon _{e}.
\end{equation}
The realistic excess noise $\varepsilon _{r}$ , which reflects real
experiment phenomena, can be calculated by

\begin{equation}
\varepsilon _{r}=\left( \eta T\left( \varepsilon _{a}+\varepsilon
_{l}\right) +\varepsilon _{b}\right) /\eta T=\varepsilon _{a}+\varepsilon
_{l}+\varepsilon _{b}/\eta T.
\end{equation}

Although the excess noise terms $\varepsilon _{a}$, $\varepsilon _{l}$, $
\varepsilon _{b}$ are constant, the calculated realistic excess noise will
increase when the transmission efficiency decreases. For example, a CV QKD
system using the GG02 protocol has an excess noise of 0.01 when the total
transmission efficiency $\eta T$ is 0.1 (39 km), $\varepsilon
_{a}+\varepsilon _{l}=0.005$, and $\varepsilon _{b}/\eta T=0.005$. Then, the
realistic excess noise (green dash-dot line) and the realistic rate-distance
limits (blue solid line) are shown in Fig. 3.

% FIG. 3
\begin{figure}[tbp]
\centerline{
\includegraphics[width=86mm]{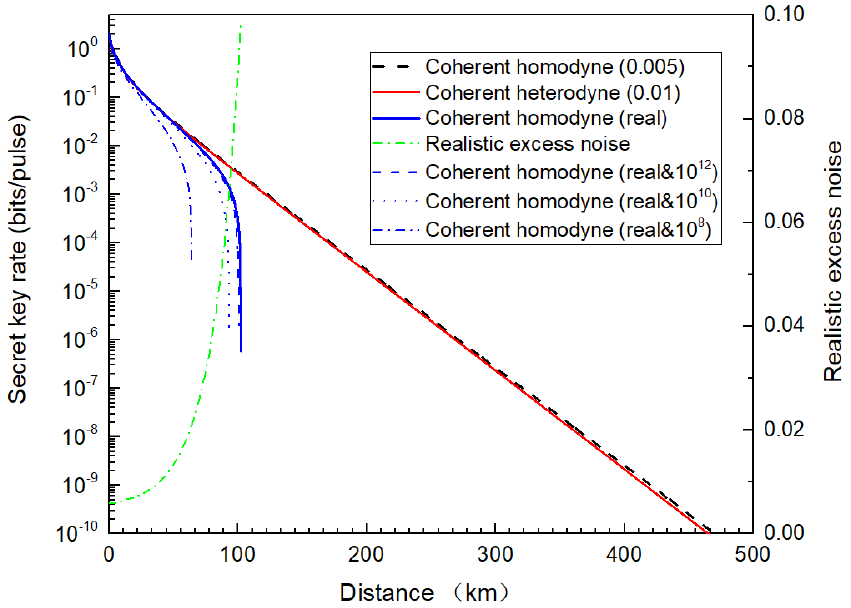}
} \vspace{0.05in} %\setlength{\columnwidth}{3.2in}
%\centerline{
%}
\caption{ Realistic rate-distance limits and the improved rate-distance limit.}
\label{Fig 3}
\end{figure}

In Fig. 3, although a high efficiency of $\beta =0.99$ is used, because the realistic excess noise rapidly increases with the transmission distance, the maximum transmission distance is limited to about 100 km. The blue dashed line, blue dotted line, and blue dash-dot line are the conditions considering
the finite-size effect which will be discussed in detail later. As Eve cannot
access Bob's setup, if we can calibrate $\varepsilon _{b}$ accurately and
treat it similar to the electronic noise, the rate-distance limit in
the asymptotic case will be represented by the black dashed line in Fig. 3 for pure
excess noise, i.e., the condition $\varepsilon _{p}=\varepsilon
_{a}+\varepsilon _{l}=0.005$. The red solid line represents the secret key
rate versus the distance with the ideal excess noise of $\varepsilon =0.01$.
Thus, we can see that the measurement of $\varepsilon _{b}$ is crucial for CV QKD to break the realistic rate-distance limit; meanwhile, we assume that it is also very challenging to achieve
experimentally.

\textit{Rate-distance limit considering the finite-size effect.}---In the above analysis, we mainly talk about the rate-distance limits in the asymptotic case. As the total number of data samples in experiments is finite, the finite-size effect must be considered. Thanks to the pioneering
work of Leverrier \cite{Leverrier2}, the finite-size effect has been
explored and several important studies were subsequently performed \cite{PWang,Papanastasiou,XYZhang}.
Typically, the expression used to calculate the secret key rate considering the finite-size effect is

\begin{equation}
\Delta I_{AB}^{f}=\left( n/N\right) \cdot \left( \beta
I_{AB}-S\left( y:E,\delta _{PE}\right) -\Delta \left( n,\delta \right)
\right) .
\end{equation}

where $n$ is number of data samples used to distill the secret key rate,
$m$ is number of the data samples used for parameter estimation, and $N=m+n$
represents the total number of samples. $I_{AB}$ indicates the Shannon
mutual information between Alice and Bob. $S\left( y:E,\delta _{PE}\right) $ represents the
maximum of the Holevo information compatible with the statistics, except
with the probability $\delta _{PE}$. $\Delta \left( n,\delta \right) $ is a
correction term for the achievable mutual information in the finite case, and
$\delta $ is the probability of an error during privacy amplification. Usually,
conservative values of $\delta _{PE}=\delta =10^{-10}$ are utilized.

% FIG. 4
\begin{figure}[tbp]
\centerline{
\includegraphics[width=86mm]{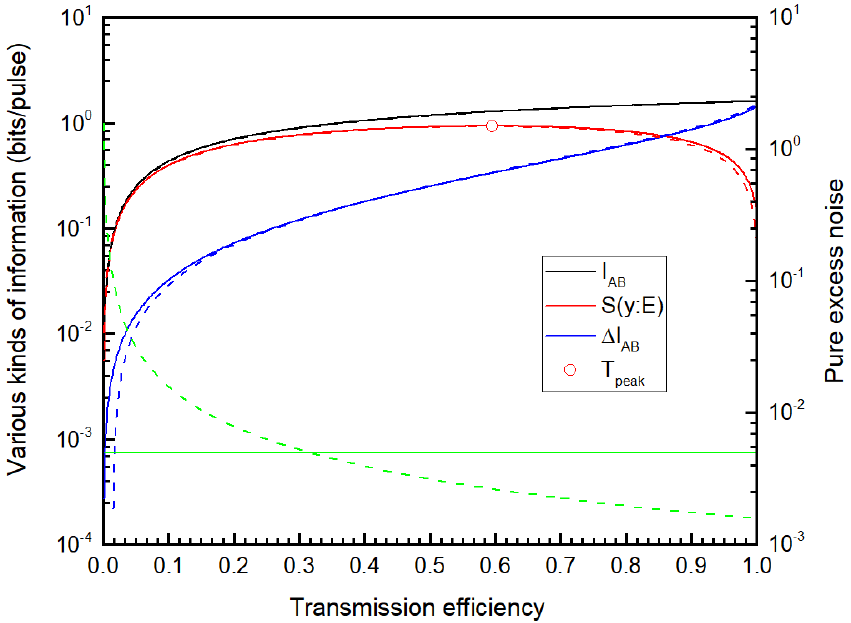}
} \vspace{0.05in} %\setlength{\columnwidth}{3.2in}
%\centerline{
%}
\caption{ Various kinds of information versus the transmission efficiency.}
\label{Fig 4}
\end{figure}

The results of numerical calculations are presented in Fig. 4. The blue,
black, and red solid curves represent the secret key rate $\Delta I_{AB}$, Shannon mutual information $I_{AB}$, and the Holevo bound information $S\left( y:E\right)$ versus $T$ respectively. The three solid curves are based on the condition that the pure
excess noise $\varepsilon _{p}=0.005$\ has a constant value (green solid
line). All dashed curves represent the corresponding information versus $T$ under
the condition that the variable $\sigma ^{2}\ $ is constant. The black dashed line overlaps the black solid line completely. In either case, the curve that represents $S\left( y:E\right)$ versus $T$ is a
convex curve. For a constant $\sigma ^{2}\ $, the pure excess noise
(green dashed line) varies with the transmission efficiency, which is
contradictory to real-life conditions. Thus, the condition that the pure excess
noise is a constant value should be considered. The transmission
efficiency that corresponds to the extreme point (red circle) of the red
solid curve is denoted as $T_{peak}$, and its square root is $t_{peak}$.
This means that calculating the maximum information $S\left( y:E,\delta
_{PE}\right)$ should be divided into two situations (different from \cite%
{Leverrier2}). When $t<t_{peak}$, the value $t_{\max }$ should be used to
calculate $S\left( y:E,\delta _{PE}\right) $. Here, $t_{\max }$ and the following $t_{\min }$ are the bounds of the confidence region of the estimated variable $\hat{t}$. At every point of this part
red solid curve, the dependence of $S\left( y:E\right)$ on the
variable $t$ can be represented as

\begin{equation}
{{\left. \frac{\partial S\left( y:E \right)}{\partial t} \right|}_{{{\sigma }^{2}}}}>0
\end{equation}

It is noted that at every point of red solid line, the constant pure
excess noise corresponds to a constant $\sigma ^{2}$; thus Eq. (7) can
be established around a fixed transmission distance. When $t>t_{peak}$, $t_{\min }$ should be used to calculate $S\left( y:E,\delta_{PE}\right)$. Thus, tighter rate-distance limits can be achieved, as
shown in Fig. 5. From left to right, the total numbers of samples are $10^{8}$
(blue lines), $10^{10}$ (green lines), and $10^{12}$ (red lines). The
dashed curves are loose limits, and the solid curves are tight limits. When
the total number of samples is larger, the difference between the solid and dashed lines of the same color is minimized. The black line represents the asymmetrical case. It is noted that the solid lines
only reflect the expected case $E\left(\hat{t}\right)$ or $E\left(\hat{\sigma}\right)$. The estimated value, $\hat{t}$\ or $\hat{\sigma}^{2}$, may
be any value in the confidence region $\left[ t_{\min },t_{\max }\right] $ or $\left[ \sigma _{\min }^{2},\sigma _{\max }^{2}\right]$ with a probability of $1-\delta _{PE}$. Thus, the rate-distance limit should be distributed in a region among the red dotted line ($L_{min}$ bound) and red dot-dash line ($L_{max}$ bound) when the total number of samples is of the order of $10^{12}$. The longest transmission distance is nearly 200 km. In an experiment, processing such a large number of data samples is very challenging and collecting these samples also requires the system to be very stable.

% FIG. 5
\begin{figure}[tbp]
\centerline{
\includegraphics[width=86mm]{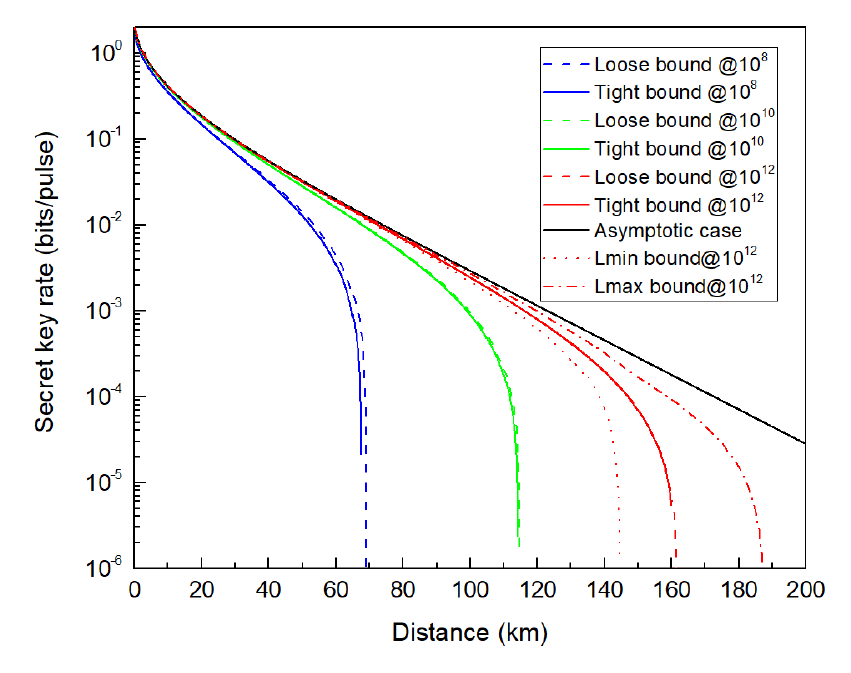}
} \vspace{0.05in} %\setlength{\columnwidth}{3.2in}
%\centerline{
%}
\caption{ Tight and loose rate-distance limits under pure excess noise conditions.}
\label{Fig 5}
\end{figure}

The realistic rate-distance limits considering the finite-size effect with different total numbers of samples are shown in Fig. 3. From left to right, the total numbers of samples for the blue
dot-dash, blue dotted, and blue dashed lines are of the order of $10^{8}$, $10^{10}$, and $10^{12}$, respectively. These realistic rate-distance limits generally reflect the limitation of the CV QKD
experimental system. The above analysis was mainly focused on the GG02 protocol; however, other protocols also have similar realistic rate-distance limits.

\textit{Conclusion.}---In this paper, we study the realistic rate-distance limits of CV QKD. At
first, we present the rate-distance limits and the experimental results of
several typical protocols. Theoretical calculations indicate that the
longest distance of 500 km can be achieved. However, the real transmission
distance is limited to $\sim $100 km. Based on the law that the excess noise
increases with the transmission distance, we find that the excess noise
generated on Bob's side and the method for calculating of the excess noise
are the primary factors behind such a relationship. Thus, realistic
rate-distance transmission limits are presented. The method for breaking the
realistic rate-distance limits is to calibrate the excess noise on Bob's
side and treat it as electronic noise. It is expected that this
challenging experimental work could be performed in the future. We also found that the improvement in the rate-distance limit due to a higher reconciliation efficiency is dependent on the excess noise. If the excess noise is larger, the improvement is more obvious. Furthermore, the rate-distance limit considering the finite-size effect is restudied, and a tighter limit is presented.

\textit{Acknowledgments.}---This research was supported by the Key Project of the Ministry of Science and Technology of China (2016YFA0301403), the National Natural Science Foundation of China (NSFC) (11504219 and 61378010), the Shanxi 1331KSC, and the Program for the Outstanding Innovative Teams of Higher Learning Institutions of Shanxi.


\begin{thebibliography}{99}

\bibitem{Scarani} V. Scarani, H. Bechmann-Pasquinucci, N. J. Cerf, M. Dusek,
N. Lutkenhaus, and M. Peev, Rev. Mod. Phys. 81, 1301 (2009).

\bibitem{Weedbrook1} C. Weedbrook, S. Pirandola, R. Garcia-Patron, N. J.
Cerf, T. C. Ralph, J. H. Shapiro, and S. Lloyd, Rev. Mod. Phys. 84, 621
(2012).

\bibitem{Pirandola} S. Pirandola, R. Laurenza, C. Ottaviani, and L. Banchi.
Nat. Commun. 8, 15043 (2017).

\bibitem{Ralph} T. C. Ralph, Phys. Rev. A 61, 010303 (1999).

\bibitem{Cerf} N. J. Cerf and M. Levy,and G. VanAssche, Phys. Rev. A 63, 052311 (2001).

\bibitem{Gottesman} D. Gottesman and J. Preskill, Phys. Rev. A 63, 022309 (2001).

\bibitem{Silberhorn} C. Silberhorn, N. Korolkova, and G. Leuchs, Phys. Rev. Lett. 88, 167902 (2002).

\bibitem{Grosshans1} F. Grosshans and P. Grangier, Phys. Rev. Lett. 88, 057902 (2002).

\bibitem{Grosshans2} F. Grosshans, G. Van Assche, J. Wenger, R. Brouri, N. J. Cerf, and P. Grangier, Nature 421, 238, (2003).

\bibitem{Weedbrook2} C. Weedbrook, A. M. Lance, W. P. Bowen, T. Symul, T. C. Ralph, and P. K. Lam, Phys. Rev. Lett. 93, 170504 (2004).

\bibitem{Garcia-Patron} R. Garcia-Patron and N. J. Cerf, Phys. Rev. Lett. 97, 190503 (2006).

\bibitem{Lodewyck1} J. Lodewyck, M. Bloch, R. Garcia-Patron, S. Fossier, E. Karpov, E. Diamanti, T. Debuisschert, N. J. Cerf, R. Tualle-Brouri, S. W. McLaughlin, and P. Grangier, Phys. Rev. A 76, 042305 (2007).

\bibitem{XYWang1} X. Y. Wang, Z. L. Bai, S. F. Wang, Y. M. Li, and K. C. Peng, Chin. Phys. Lett. 30, 010305 (2013).

\bibitem{Jouguet}  P. Jouguet, S. Kunz-Jacques, A. Leverrier, P. Grangier, and E. Diamanti, Nat. Photon. 7, 378 (2013).

\bibitem{SXLong} X. L. Su, Chin. Sci. Bull. 59, 1083 (2013).

\bibitem{Usenko} V. C. Usenko and F. Grosshans, Phys. Rev. A 92, 062337 (2015).

\bibitem{CWang} C. Wang, D. Huang, P. Huang, D. Lin, J. Peng, and G. Zeng, Sci. Rep. 5, 14607 (2015).

\bibitem{BQi} B. Qi, P. Lougovski, R. Pooser, W. Grice, and M. Bobrek, Phys. Rev. X 5, 041009 (2015).

\bibitem{DHuang1} D. Huang, P. Huang, D. Lin, and G. Zeng, Sci. Rep. 6, 19201 (2016).

\bibitem{XYWang2} X. Y. Wang, W. Y. Liu, P. Wang, and Y. M. Li, Phys. Rev. A 95, 062330 (2017).

\bibitem{Fossier} S. Fossier, E. Diamanti, T. Debuisschert, A. Villing, R. Tualle-Brouri, and P. Grangier, New J. Phys. 11, 045023 (2009).

\bibitem{DHuang2}  D. Huang, P. Huang, H. Li, T. Wang, Y. Zhou, and G. Zeng, Opt. Lett. 41, 3511 (2016).

\bibitem{YMLi}  Y. M. Li, X. Y. Wang, Z. L. Bai, W. Y. Liu, S. S. Yang, and K. C. Peng, Chin. Phys. B 26, 040303 (2017).

\bibitem{Karinou} F. Karinou , H. H. Brunner, C. F. Fung, L. C. Comandar, S. Bettelli, D. Hillerkuss, M. Kuschnerov, S. Mikroulis, D. Wang , C. Xie, M. Peev, and A. Poppe, IEEE Photonic Tech. Lett. 30, 650 (2018).

\bibitem{Milicevic} M. Milicevic, C. Feng, L. M. Zhang and P. G. Gulak, NPJ Quantum Inf. 4, 21 (2018).

\bibitem{PWang} P. Wang, X.Y. Wang, J.Q. Li, and Y.M. Li, Opt. Express 25, 27995 (2017).

\bibitem{ZQu} Z. Qu, I. B. Djordjevic, and M. A. Neifeld. 41, 5507 (2016).

\bibitem{Leverrier1} A. Leverrier and P. Grangier, Phys. Rev. A 83, 042312 (2011).

\bibitem{Pirandola2}S. Pirandola, C. Ottaviani, G. Spedalieri, C. Weedbrook, S. L. Braunstein, S. Lloyd, T. Gehring, C. S. Jacobsen, and U. L. Andersen, Nat. Photonics 9, 397 (2015).

\bibitem{Lodewyck2} J. Lodewyck, T. Debuisschert, R. Tualle-Brouri, and P. Grangier, Phys. Rev. A 72, 050303(R) (2005).

\bibitem{Leverrier2} A. Leverrier, F. Grosshans, and P. Grangier, Phys. Rev. A 81, 062343 (2010).

\bibitem{Papanastasiou} P. Papanastasiou, C. Ottaviani, and S. Pirandola, Phys. Rev A 96, 042332 (2017).

\bibitem{XYZhang} X. Zhang, Y. Zhang, Y. Zhao, X. Wang, S. Yu, and H. Guo, Phys. Rev A 96, 042334 (2017).

\end{thebibliography}
\end{document}